\documentclass[twocolumn,amsmath,showpacs,amsfonts,aps,prc,floatfix]{revtex4}
\usepackage{graphicx}
\usepackage{bm}

\begin{document}

\title{Hydrodynamical analysis of centrality dependence of charged particle's multiplicity  
in $\sqrt{s}_{NN}$=2.76 TeV Pb+Pb collisions}
\author{A. K. Chaudhuri}
\email[E-mail:]{akc@veccal.ernet.in}
\author{Victor Roy}
\email[E-mail:]{victor@veccal.ernet.in}
\affiliation{Variable Energy Cyclotron Centre, 1/AF, Bidhan Nagar, 
Kolkata 700~064, India}

\begin{abstract}

In Israel-Stewart's theory of dissipative hydrodynamics, we have analyzed the recent ALICE data for the centrality dependence of charged particle multiplicity,
centrality dependence of integrated elliptic flow and 0-5\% charged particles $p_T$ spectra in $\sqrt{s}_{NN}$=2.76 TeV Pb+Pb collisions and determined the initial or the thermalisation time.  
Analysis indicate that the ALICE data disfavor very early thermalsiation $\tau_i$=0.2 fm, or very late thermalsation $\tau_i$= 4fm. Data are best explained   for thermalisation time $\tau_i=0.88^{-0.14}_{+0.68}$ fm.
\end{abstract}

\pacs{47.75.+f, 25.75.-q, 25.75.Ld} 

\date{\today}  

\maketitle 
 
Lattice simulations of QCD indicate that the strongly interacting nuclear matter can under go a confinement-deconfinement cross-over transition  \cite{Karsch:2008fe,Cheng:2009zi,Aoki:2006we,Fodor:2010zz}. Experiments at Relativistic Heavy Ion Collider (RHIC), produced convincing evidences that in $\sqrt{s}_{NN}$=200 GeV Au+Au collisions a {\em collective} QCD medium is created \cite{BRAHMSwhitepaper,PHOBOSwhitepaper,PHENIXwhitepaper,STARwhitepaper}, though it is uncertain whether or not the matter produced can be characterised as 
Quark-Gluon Plasma (QGP), the lattice QCD predicted deconfined phase.
The issue is expected to be settled at Large Hadron Collider, where   Lead nuclei will collide head on at enormous energy $\sqrt{s}_{NN}$=5.5 TeV.  Recently, ALICE collaboration \cite{Aamodt:2010pb,Collaboration:2010cz} measured the centrality dependence of  charged particle multiplicity in $\sqrt{s}_{NN}$=2.76 TeV Pb+Pb collision. The centrality dependence is similar to that obtained in $\sqrt{s}_{NN}$=200 GeV Au+Au collisions at RHIC, charged particle multiplicity, normalized by the participant nucleon pair increase by a factor of 2 from peripheral (70-80\%) to central (0-5\%) collisions. ALICE collaboration also published charged particles $p_T$ spectra and elliptic flow \cite{Aamodt:2010jd,Aamodt:2010pa}. In 0-5\% collisions, $p_T$ spectra of charged particle's are suppressed by a factor $R_{AA}\sim 0.14$ at $p_T$=6-7 GeV, which is smaller than at lower energies. In  peripheral collision, suppression is modest, $R_{AA}\approx$0.6-0.7. Integrated elliptic flow   is $\sim$30\% more than that in Au+Au   collisions at RHIC.

Hydrodynamic models have been used extensively to analyze the experimental data in $\sqrt{s}_{NN}$=200 GeV Au+Au collisions at RHIC and obtain information about the initial condition of the produced medium. 
Hydrodynamic models require the   assumption of 'local' thermal equilibrium.
If the assumption is met, relativistic hydrodynamic equations can be solved to trace-back to the initial fluid condition from experimental data.  
At RHIC energy, near ideal QGP fluid, initialized to central energy density $\varepsilon_i\approx$ 30 $GeV/fm^3$ at initial time $\tau_i\approx$0.6 fm, explains a large variety of experimental data \cite{QGP3,Luzum:2008cw,Chaudhuri:2008sj,Chaudhuri:2008ed,Chaudhuri:2009uk}. 
 
One of the important issue in a hydrodynamical model analysis is the initial or the thermalisation time.  In the present letter, we have analysed the 
  ALICE data  for the centrality dependence of charged particle  multiplicity \cite{Collaboration:2010cz}, centrality dependence of integrated elliptic flow \cite{Aamodt:2010pa} and charged particles $p_T$ spectra in 0-5\% collision \cite{Aamodt:2010jd}, to determine the thermalisation time in LHC energy collisions. The three data sets are best explained if thermalisation time is $\tau_i\approx$1 fm.  Small thermalisation time $\tau_i$=0.2 fm, or large thermalisation time $\tau_i$=4 fm are disfavored by the data. 

We assume that in $\sqrt{s}_{NN}$=2.76 TeV Pb+Pb collisions, a baryon free QGP fluid is formed.  Viscosity to entropy ratio of the fluid is assumed to be   $\eta/s=1/4\pi$ \cite{Policastro:2001yc,Kovtun:2003wp}. 
The space-time evolution of the fluid is obtained by solving Israel-Stewart's 2nd order theory \cite{IS79,Heinz:2005bw},
 
\begin{eqnarray}  
\partial_\mu T^{\mu\nu} & = & 0,  \label{eq1} \\
D\pi^{\mu\nu} & = & -\frac{1}{\tau_\pi} (\pi^{\mu\nu}-2\eta \nabla^{<\mu} u^{\nu>}) \nonumber \\
&-&[u^\mu\pi^{\nu\lambda}+u^\nu\pi^{\nu\lambda}]Du_\lambda. \label{eq2}
\end{eqnarray}

Eq.\ref{eq1} is the conservation equation for the energy-momentum tensor, $T^{\mu\nu}=(\varepsilon+p)u^\mu u^\nu - pg^{\mu\nu}+\pi^{\mu\nu}$, 
$\varepsilon$, $p$ and $u$ being the energy density, pressure and fluid velocity respectively. $\pi^{\mu\nu}$ is the shear stress tensor. Eq.\ref{eq2} is the relaxation equation for the shear stress tensor $\pi^{\mu\nu}$.   
In Eq.\ref{eq2}, $D=u^\mu \partial_\mu$ is the convective time derivative, $\nabla^{<\mu} u^{\nu>}= \frac{1}{2}(\nabla^\mu u^\nu + \nabla^\nu u^\mu)-\frac{1}{3}  
(\partial . u) (g^{\mu\nu}-u^\mu u^\nu)$ is a symmetric traceless tensor. $\eta$ is the shear viscosity and $\tau_\pi$ is the relaxation time.  It may be mentioned that in a conformally symmetric fluid relaxation equation can contain additional terms  \cite{Song:2008si}.    
Assuming boost-invariance, Eqs.\ref{eq1} and \ref{eq2}  are solved in $(\tau=\sqrt{t^2-z^2},x,y,\eta_s=\frac{1}{2}\ln\frac{t+z}{t-z})$ coordinates, with the code 
  "`AZHYDRO-KOLKATA"', developed at the Cyclotron Centre, Kolkata.
 Details of the code can be found in \cite{Chaudhuri:2008sj}. 
 
Hydrodynamic equations  are closed with an equation of state $p=p(\varepsilon)$.
In the present study, we use an equation of state where the Wuppertal-Budapest \cite{Aoki:2006we}
lattice simulations for the deconfined phase is smoothly joined at $T=T_c=174$ MeV, with hadronic resonance gas EoS comprising all the resonances below mass $m_{res}$=2.5 GeV.
 
 \begin{figure}[t]
\center
 \resizebox{0.30\textwidth}{!}{%
  \includegraphics{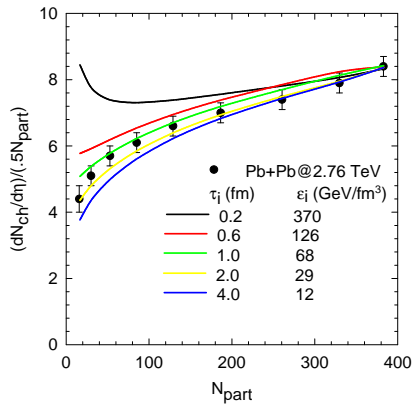}
}
\caption{(color online) The black circles are ALICE data for centrality dependence of charged particle's multiplicity per participant pair in   $\sqrt{s}_{NN}$= 2.76 GeV Pb+Pb collisions \cite{Collaboration:2010cz}.
The black, red, green,yellow and blue lines are hydrodynamic simulations  with 
initial time $\tau_i$=0.2, 0.6, 1.0, 2.0 and 4.0 fm  respectively.}
  \label{F1}
\end{figure}  
 
Solution of partial differential equations (Eqs.\ref{eq1},\ref{eq2}) requires initial conditions, e.g.  transverse profile of the energy density ($\varepsilon(x,y)$), fluid velocity ($v_x(x,y),v_y(x,y)$) and shear stress tensor ($\pi^{\mu\nu}(x,y)$) at the initial time $\tau_i$. One also need to specify the viscosity ($\eta$) and the relaxation time ($\tau_\pi$). A freeze-out prescription is also needed to convert the information about fluid energy density and velocity to particle spectra and compare with experiment. We assume that in an impact parameter ${\bf b}$ collision, at the initial time $\tau_i$, initial energy density is   distributed as in a Glauber model \cite{QGP3},

\begin{equation} \label{eq3}
\varepsilon({\bf b},x,y)=\varepsilon_i[(1-f) N_{part}({\bf b},x,y) + f N_{coll}({\bf b},x,y)],
\end{equation}

 $N_{part}$ and $N_{coll}$ in Eq.\ref{eq3} are the transverse profile of the average number of  participants and average number collisions respectively.  $f$ is the hard scattering fraction. We assume that $\sqrt{s}_{NN}$=2.76 TeV Pb+Pb   collisions are dominated by hard scattering and fix $f=0.9$. We also assume that at the initial time fluid velocity is zero. We initialize the shear stress tensor to boost-invariant value, $\pi^{xx}=\pi^{yy}=2\eta/3\tau_i$, $\pi^{xy}$=0.
For the relaxation time, we use the   Boltzmann estimate $\tau_\pi=3\eta/2p$. 
Fluid viscosity is assumed to be $\eta/s=1/4\pi$ and we neglect any temperature dependence of $\eta/s$.
Hydrodynamics also require a freeze-out condition. We assume that the fluid freeze-out at a fixed temperature $T_F$=130 MeV. 
In viscous hydrodynamics, particle production has contributions from the non-equilibrium part of the distribution function. We include the non-equilibrium contribution.  
We note that the freeze-out condition does not account for the validity condition for viscous hydrodynamics, i.e. relaxation time for dissipative fluxes are much greater than the inverse of the expansion rate, $\tau_R \partial_\mu u^\mu << 1$. Recently, Dusling and Teaney \cite{Dusling:2007gi} implemented dynamical freeze-out condition. 
In dynamical freeze-out, non-equilibrium effects are stronger than obtained at fixed temperature freeze-out. Freeze-out at $T_F$=130 MeV however satisfies the viscous hydrodynamic condition that non-equilibrium contribution to particle production is much smaller than the equilibrium contribution. 

 \begin{table*}[t]
\caption{\label{table1} Central energy density ($\varepsilon_i$) and temperature ($T_i$) required to reproduce experimental charged particle's multiplicity in 0-5\% Pb+Pb collisions, for   $\tau_i$=0.2-4.0 fm. Freeze-out temperature $T_F$=130 MeV. Model estimates for the charged particle multiplicity is also noted. In the last 4 columns, $\chi^2/N$ for the data sets analysed are noted. }
\begin{ruledtabular} 
  \begin{tabular}{|c|c|c|c|c|c|c|c|}\hline
$\tau_i$ &  $\varepsilon_i$ & $T_i$ & $\left (\frac{dN_{ch}}{dy} \right )_{TH}$
& $\chi^2/N$ \text{for} & $\chi^2/N$ \text{for} & $\chi^2/N$ \text{for}\footnote{data for 0-5\% $p_T$ spectra are read from \cite{Aamodt:2010jd}. We assume 7\% error in the data.}
& $\chi^2/N$ \text{for}\\  
(fm)      & $(GeV/fm^{3})$  & (MeV) &  &$\frac{1}{.5N_{part}}\frac{dN_{ch}}{dy}$  & $v_2$ & 0-5\% $p_T$\text{-spectra} &$\frac{1}{.5N_{part}}\frac{dN_{ch}}{dy}+v_2+p_T$\text{-spectra} \\ \hline
0.2  &  $370\pm 30$ & $690\pm 10$ &  1611 &29.0&4.7&12.2 &14.5  \\ \hline
0.6  &  $126\pm 9$ & $532\pm9$    & 1599&4.1 &5.6 &5.2&5.0 \\ \hline
1.0  &   $72\pm 5$ & $464\pm7$  &  1614 &0.9 &6.2 &2.6 &2.9\\ \hline
2.0  &   $29\pm 1$ & $367\pm 5$  &1603 &0.3 &11.8 &3.2 & 4.2\\\hline
4.0  &   $12\pm 1$ & $301\pm 3$   &1604 &1.7&20.2&15.1 &13.1\\ \hline
\end{tabular}\end{ruledtabular}  
\end{table*}

The central energy density $\varepsilon_i$ at the initial time $\tau_i$ is fixed to reproduce experimental charged particle's multiplicity $\frac{dN_{ch}}{dy}=1601\pm 60$ in 0-5\% Pb+Pb collision. 
  For initial time $\tau_i$=0.2, 0.6, 1.0, 2.0 and 4.0 fm, we have simulated 0-5\% Pb+Pb collisions and computed negative pion multiplicity. Resonance production is included. Noting that    pion's constitute $\approx$85\% of the total charged particles, $\pi^-$ multiplicity is multiplied by the factor $2\times 1.15$ to compare with experimental charged particle multiplicity.
  Irrespective of the initial time,   central energy density can be varied  to reproduce
experimental multiplicity in 0-5\% collision centrality. In table.\ref{table1}, central energy density and temperature required to reproduce experimental charged particle multiplicity in 0-5\% collisions are listed. Uncertainty in $\varepsilon_i$ or $T_i$ reflect the   uncertainty
 in ALICE measurements. 
If QGP fluid is thermalised in the time scale $\tau_i$=0.2 fm, very large energy density $\sim$ 370 $GeV/fm^3$ is required to reproduce the experimental multiplicity. Energy density is reduced if thermalisation time increased. One also note that for comparable thermalisation time, at LHC, fluid is produced at much higher energy density or temperature than at RHIC energy collisions. For example, for $\tau_i$=0.6 fm, $\varepsilon_i\approx$ 126  $GeV/fm^3$ at LHC energy collisions, as compared to  $\varepsilon_i\approx$ 30  $GeV/fm^3$ at RHIC collisions.

    \begin{figure}[t]
\center
 \resizebox{0.30\textwidth}{!}{%
  \includegraphics{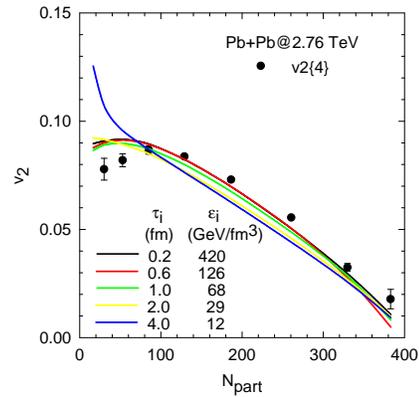}
}
\caption{(color online) Black, red, green, yellow and blue lines are hydrodynamical model simulations for elliptic flow for pions, for initial time $\tau_i$=0.2, 0.6, 1.0, 2.0 and 4.0 fm. The black circles  are  the ALICE measurement \cite{Aamodt:2010pa} for charged particle's elliptic flow in   4-particle cumulant method. 
}
  \label{F2}
\end{figure}  
 
With the initial conditions fixed, we have simulated Pb+Pb collisions over a wide range of collision centralities. In Fig.\ref{F1} simulation results are compared with  the ALICE measurements for the centrality dependence of charged particle multiplicity per participant nucleon pair ($\frac{1}{.5N_{part}}\frac{dN_{ch}}{dy}$). ALICE measurements are not reproduced if the fluid thermalises early, $\tau_i$=0.2 fm, multiplicity in peripheral
collisions are over predicted. Data are better reproduced for $\tau_i >$0.2 fm.
Very different trend of the simulated $\frac{1}{.5N_{part}}\frac{dN_{ch}}{dy}$  for $\tau_i$=0.2 fm is puzzling. In a Glauber model of initialisation, one expects that with different initial times, centrality dependence will be qualitatively similar, though differ in quantitative magnitude. Indeed, simulation results do indicate that for initial time $\tau_i >$0.2 fm, though quantitatively different, qualitatively, centrality dependence of multiplicity per participant pair is similar. For $\tau_i$=0.2, qualitative nature is changed. 
The reason can be understood as follows:
 For $\tau_i$=0.2 fm, initial pressure, even in a peripheral collision, is very high. Due to large pressure gradient fluid can accelerates rapidly and achieve large velocity. We have checked that for $\tau_i$=0.2 fm, in central and peripheral collisions, at the freeze-out fluid velocity ($v_r$) can be as high as $v_r$=0.8c. In contrast, for larger thermalisation time $\tau_i$=.6 fm, 
though in a central collisions, freeze-out velocity is $v_r \sim$0.8c, in a peripheral collision, $v_r\sim$0.6c. Even though initialisation with $\tau_i$=0.2 and 0.6 fm, both   produces similar number of particles in a central collisions, in peripheral collisions, initialisation with $\tau_i$=0.2 fm produces more particles due to increased fluid velocity.   We may note here that at very early time, e.g. $\tau_i$=0.2 fm, non-equilibrium effect can be large and even viscous hydrodynamics may not be applicable. Failure to explain the data with initial time $\tau_i$=0.2 fm may be  either due to improper choice of initial time, or due to inapplicability of viscous hydrodynamics or both. But charged particles multiplicity data definitely  disfavor very early initialisation.
 
Integrated elliptic flow is an important observable in relativistic energy collisions. It is a measure of collectivity in the medium. In Fig.\ref{F2}, centrality dependence of simulated (integrated) elliptic flow in $\sqrt{s}_{NN}$=2.76 TeV Pb+Pb collisions are compared with ALICE measurements \cite{Aamodt:2010pa}. Even though $\tau_i$=0.2 fm is not consistent with centrality dependence of charged particle multiplicity, simulated flows agree with the experiment. Indeed, for $\tau_i$=0.2-1.0 fm, simulated flows are nearly identical (differ by less than 3\%) and  well explains the ALICE data. For higher thermalisation time $\tau_i >$ 2 fm, description to the data gets poorer. Also $v_2$ in peripheral collisions continue to increase as oppose  to the experimental data. Naively, one expects that for late initialsiation, pressure will be low and expansion time will shorten and elliptic flow will not grow. Explicit simulation indicate that for late thermalisation, though the fluid pressure is low, expansion time is not shortened likewise. For example, for initial time $\tau_i$=0.6 fm, in a peripheral collision, (central) fluid   freezes out at $\tau_F\approx$6.6 fm  
 and for $\tau_i$=4.0 fm,  expansion time is shortened marginally, fluid freeze-out at,   $\tau_F\approx$9.9 fm. However, growth of momentum anisotropy, 
 $\varepsilon_p=\frac{\int dxdy(T^{xx}-T^{yy})}{\int dxdy(T^{xx}+T^{yy})}$
 differ markedly. For $\tau_i$=0.6 fm, $\varepsilon_p$, after few fm of evolution saturate to $\varepsilon_p\sim$0.13.
Due to decreased pressure, for late initialisation, momentum anisotropy grow rather slowly. It do not saturate, but continue to grow till freeze-out and at freeze-out $\varepsilon_p\sim$0.17. Enhanced momentum anisotropy for late thermalisation is reflected as enhanced $v_2$.  

\begin{figure}[t]
\center
 \resizebox{0.30\textwidth}{!}{%
  \includegraphics{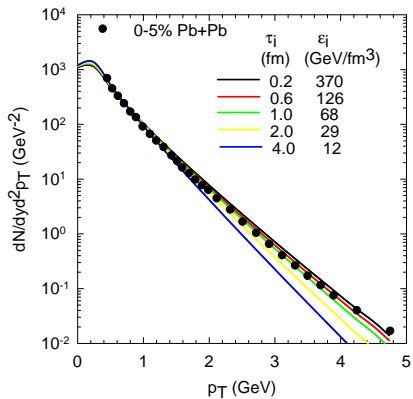}
}
\caption{(color online) black   circles are charged particle's spectra in 0-5\%   Pb+Pb collisions at  $\sqrt{s}_{NN}$=2.76 TeV \cite{Aamodt:2010jd}. The black, red, green, yellow and blue lines are hydrodynamic model simulations for  charged particle's spectra for initial time $\tau_i$=0.2, 0.6, 1.0, 2.0 and 4.0 fm respectively.}\label{F3}
\end{figure}

 In Fig.\ref{F3} ALICE measurements   \cite{Aamodt:2010jd} for the charged particle's  spectra in 0-5\%   collision are compared with hydrodynamical simulations. 
Slope of the spectra is enhanced for smaller initial time. 
It is expected also.  Other conditions remaining unchanged, fluid initialized at higher temperature produces more high $p_T$ particles than fluid initialized at lower temperature. 0-5\% spectra are reasonably well explained in simulations with thermalisation time $\tau_i$=0.6-2.0 fm.  Description to the data gets poorer for   thermalisation time $\tau_i >$ 2 fm.

For a quantitative analysis, we have computed  $\chi^2/N$ values for the data sets analysed. They are noted in table.\ref{table1}. 
From the $\chi^2$ values, it is apparent that if only a single data set is considered, one may largely underestimate or overestimate   the thermalisation time. For example, if one consider only the   elliptic flow data, one may conclude  that the thermalisation time at LHC is $\tau_i$=0.2 fm. However, thermalisation time $\tau_i$=0.2 fm is not prefered by the ALICE data on the centrality dependence of the charged particles multiplicity. Perfect fit to the multiplicity data is obtained for thermalisation time 
$\tau_i$=1.0-2.0 fm.   0-5\% $p_T$ spetcra in 0-5\% collisions on the otherhand  
prefer thermalisation time $\tau_i$=1.0 fm. 
Thermalisation time at LHC energy should be obtained by analysing all the data sets simultaneously. 
In the last column of table.\ref{table1}, we have noted the $\chi^2/N$ values for the three data sets combined. 
The combined data are best expalined for thermalisation time $\tau_i$=1 fm,  $(\chi^2/N)_{min}$=2.8. Very small thermalisation time $\tau_i$=0.2 fm or very large thermalisation time $\tau_i$=4.0 fm is not favored by the data, $\chi^2/N$ increase by a factor of $\sim$4-5 from the minimum $\chi^2/N$ value. 
Using a parabolic fit to the $\chi^2/N$ values, we could estimate the thermsalisation time rather accurately, $\tau_i=0.88^{-0.14}_{+0.68}$ fm. 
It must be mentioned that the present estimate   is obtained with a specific set of initial conditions, e.g. initial zero fluid velocity, hard scattering fraction f=0.9, boost invariant values for the initial shear stress tensors etc. We have also assumed a fixed viscosity to entropy ratio $\eta/s$=0.08.  All the possible initial conditions are not explored. If uncertinty over the initial conditions are included, the present estimate of thermalisation time at LHC energy   $\tau_i$=0.74-1.56 fm, will be even more uncertain.

In the present analysis, we have neglected bulk viscosity. In general bulk viscosity is much smaller than shear viscosity. In QCD, near the transition point, bulk viscosity can be large \cite{Kharzeev:2007wb,Karsch:2007jc}. Effect
of bulk viscosity on particle spectra and elliptic flow has been studied in \cite{Song:2008si}.  Compared to shear viscosity, effect of bulk viscosity is small. Present result that ALICE data  disfavor small or large thermalisation time, or the present estimate   of the thermalisation time 
$\tau_i=0.88^{-0.14}_{+0.68}$ fm will also remain largely unaltered even if bulk viscous effects are included.

To conclude,  analysing the  recent ALICE data for the centrality dependence of charged particle multiplicity,    integrated elliptic flow and   $p_T$ spectra, we have determined the thermalisation time for QGP fluid in   $\sqrt{s}_{NN}$=2.76 TeV Pb+Pb collisions. The combined data are best explained for thermalisation time $\tau_i=0.88^{-0.14}_{+0.68}$ fm.   Small thermalisation time $\tau_i$=0.2 fm, or large thermalisation time $\tau_i $= 4 fm, is not favored by the data.

\end{document}